\documentclass[aps,prl,reprint,superscriptaddress,amsmath,amssymb,floatfix]{revtex4-2}
\usepackage{graphicx,bm,xcolor,microtype}
\usepackage[hypertexnames=false,colorlinks=true,allcolors=blue]{hyperref}
\graphicspath{{figures/}}
\newcommand{\ii}{\mathrm i}

\newcommand{\up}{\uparrow}
\newcommand{\dn}{\downarrow}

\newcommand{\DoU}{\mathrm{DoU}}
\begin{document}
% CHECK: The title word "Complete" must be assessed against first-order control at a fixed nonzero momentum in a lossless structure.
\title{Chiral Unidirectional Resonances Enabled by Complete Coherent Radiation Control}
\author{Keren Wang}
\affiliation{College of Physics, Sichuan University, Chengdu 610064, China}
\author{Wei Wang}
\email[Corresponding author: ]{w.wang@scu.edu.cn}
\affiliation{College of Physics, Sichuan University, Chengdu 610064, China}
% PROVISIONAL working draft for author review, 2026-10-24. See _notes/LOCAL-WRITEBACK-HANDOFF.md.
\begin{abstract}
Optical resonances are commonly characterized by their quality factors, while their radiation exhibits richer properties associated with polarization and directional selectivity. Chiral and unidirectional guided resonances exploit complementary aspects of radiation, motivating a unified framework for controlling polarization and directionality together. Here we develop such a framework and introduce chiral unidirectional resonances (CURs) with arbitrarily high radiative quality factors and near-unity single-channel selectivity. We further extend this theory to guide the geometric design of arbitrary coherent emission states with prescribed radiative lifetimes. In addition, we establish scaling laws for the lifetime and radiation-state fidelity, and demonstrate robust chiral unidirectional emission over a finite momentum region. Our demonstrations provide a basis for single-sided circularly polarized light sources and polarization-matched optical coupling.
\end{abstract}
\maketitle

\typeout{CHIRALCOL=\the\columnwidth; CHIRALTEXT=\the\textwidth}

Controlling the radiation of optical resonances is central to shaping how light is stored and emitted in open photonic structures. Coupling to the surrounding continuum determines both the radiative lifetime and the polarization and directionality of the emitted field. Such control enables spectrally and spatially coherent thermal emission and structured thermal fields~\cite{wangthermal2025,sunstructured2026,sun2025}. Bound states in the continuum (BICs) suppress radiative leakage through symmetry or destructive interference~\cite{friedrich1985,marinica2008,hsu2013,hsu2016,huangreview2023}. Structural perturbations reopen the radiation channels, with their strength controlling the quality factor~\cite{rybin2017,koshelev2018,overvigselection2020}. The radiation zeros also admit a topological description: they are centers of far-field polarization vortices whose winding in momentum space defines a topological charge~\cite{zhen2014,doeleman2018}. This view connects the suppressed radiation at a BIC to the polarization of the surrounding leaky resonances and extends to the merging dynamics of higher-order topological charges~\cite{wangmerging2026}. Structural modulation controls the relative amplitudes and phases of this radiation, allowing its polarization to be tailored~\cite{liu2019,guo2017}.

A natural use of this polarization freedom is to select the handedness of the emitted light. Chiral quasi-BICs realize this selection by favoring one helicity~\cite{gorkunov2020,overvig2021,chen2023,shi2022,suncontinuum2026}, enabling circularly polarized emission, including thermal radiation~\cite{sunthermal2025}. Complementing these advances in polarization control, unidirectional guided resonances (UGRs) suppress radiation into one half-space through destructive interference while preserving emission into the other~\cite{yin2020,yin2023,choi2025,zhangphase2021}, enabling single-sided thermal emission without a metallic back-reflector~\cite{wangunithermal2026}. Both polarization and directionality are determined by complex radiation amplitudes~\cite{kang2022}; symmetry, reciprocity, and the nonresonant background constrain their relative phases~\cite{hsu2017,zhou2016}. Combining helicity selection and single-sided radiation in a single resonance therefore calls for a unified framework that retains these phases together with channel powers.

\begin{figure}[!t]
\centering\includegraphics[width=\columnwidth]{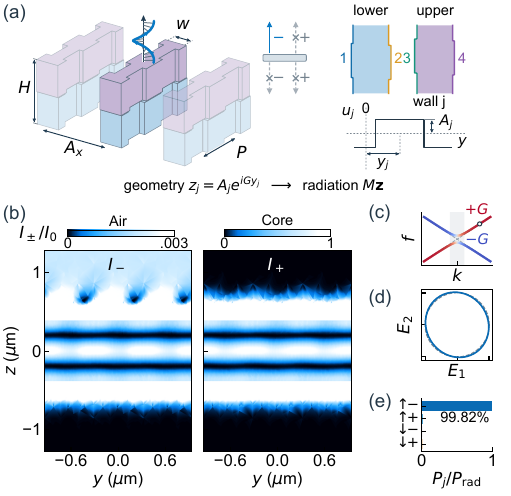}
\caption{A high-$Q$ CUR with inversion-paired walls. (a) Rails, wall masks, and a four-channel schematic with the selected channel in blue and suppressed channels in gray. Depth-region colors match the right-hand cross sections; displacement is exaggerated fivefold. (b) Eigenmode projections onto the fixed upper-outgoing circular analyzers. Core and air share a palette, with the indicated linear ranges common to both analyzers. (c) Schematic $\pm G$ parent mixing; color denotes parent content. The circle marks the design point outside the mixing region. (d) FEM upper-side Jones ellipse; the dashed circle is an ideal polarization reference. (e) Four-channel powers. Arrows indicate propagation.}
\label{fig:realization}
\end{figure}

Here we develop such a framework and introduce chiral unidirectional resonances (CURs) with arbitrarily high radiative quality factors and near-unity single-channel selectivity. An inversion-paired structure realizes joint helicity and half-space selection within the same eigenmode. However, its nonresonant background constrains the accessible coherent states and imposes a sharp fidelity bound. We further extend the theory to guide the geometric design of arbitrary coherent emission states and radiative lifetimes, encompassing quasi-BICs, chiral resonances, UGRs, and CURs within a common description. Four independent sidewall modulations yield an invertible first-order map from a prescribed radiation state and lifetime to geometry at fixed nonzero momentum in a lossless system. In addition, we establish scaling laws for radiative lifetime and radiation-state fidelity, and demonstrate persistent channel selection beyond the design momentum.

We begin by resolving the radiation of an isolated resonance into four power-normalized channels so that polarization, half-space selection and radiative lifetime are determined by the same complex amplitudes. The channel order is $(\up-,\up+,\dn-,\dn+)$. The arrows denote the two half-spaces, helicity is defined relative to each outward wave vector, and the time convention is $e^{-\ii\omega t}$. For a lossless system, temporal coupled-mode theory gives~\cite{fan2002,fan2003,suh2004}
\begin{equation}
 \bm d=\sqrt{2\gamma_{\rm rad}}\,\bm n,
 \qquad \bm n^\dagger\bm n=1,
 \qquad Q_{\rm rad}=\frac{\omega_0}{2\gamma_{\rm rad}}.
 \label{eq:rate}
\end{equation}
Within this framework, a CUR is the single-channel limit of the radiation state $\bm n$, selecting one helicity in one half-space within the same eigenmode. For the upper negative-helicity channel, the fraction of total radiated power is
\begin{equation}
 \beta_{\up-}=\frac{P_{\up-}}{P_{\rm rad}}
 =\frac{(1+\DoU)(1-s_3^\up)}{4},
 \label{eq:joint}
\end{equation}
where $\eta=P_\up/P_\dn$, $\DoU=(\eta-1)/(\eta+1)$, and $s_3^\up=(P_{\up+}-P_{\up-})/P_\up$. This fraction quantifies helicity and half-space selection jointly. More generally, an unrestricted four-channel vector has seven real degrees of freedom after removal of an irrelevant common phase: six specify the radiation state, while the norm of $\bm d$ sets the radiative lifetime. For a normalized target $\bm t$, the fidelity $|\bm t^\dagger\bm n|^2$ includes relative-phase information and reduces to $\beta_{\up-}$ for a single-channel target.

Realizing this single-channel state requires three unwanted complex amplitudes to vanish: one for the unwanted helicity in the emitting half-space and two for both helicities in the other half-space. For an unrestricted four-channel vector, these impose six real conditions while leaving the radiative lifetime free. Symmetry can relate the cancellation conditions or exclude the target itself. The number of geometric parameters alone therefore does not determine which radiation states a structure can realize.
% CHECK: This six-real-condition count applies to an unrestricted channel vector at finite output, before inversion constraints. No universal two-plus-two count is established for the paired structure.

We realize CURs with arbitrarily high radiative quality factors by designing a dielectric-rail array with inversion-paired sidewall modulations [Fig.~\ref{fig:realization}(a)]. The equal-height upper and lower regions carry inversion-related wall patterns. Periodic modulation couples a guided parent mode to radiation, while the modulation amplitudes and longitudinal shifts control the strength and phase of the contributions from each wall. We choose these parameters so that the wall radiation reinforces the upper negative-helicity channel and nearly cancels in the other three. The design thus combines helicity selection and unidirectional radiation within a single eigenmode while retaining inversion symmetry. Full-wave calculations confirm a representative CUR at finite modulation [Fig.~\ref{fig:realization}(b,d,e)]. The eigenmode has $Q\approx1.011\times10^4$ and $s_3^\up\approx-0.9968$, with approximately 99.8\% of its radiation in the upper negative-helicity channel. Inversion symmetry maps this mode at $\bm k$ to a distinct counterpropagating partner at $-\bm k$ that radiates into the lower half-space with opposite helicity. The handedness therefore characterizes the radiation at $\bm k$. The design momentum lies outside the narrow region of strong counterpropagating-parent mixing [Fig.~\ref{fig:realization}(c)], with the geometry, momentum and independent numerical comparisons specified in the Supplemental Material (SM)~\cite{SM}.
% Source: core-paper-closeout-20261003-r1/independent-FEM-lattice.npz, Q[16,16]=10113.804926036515; FEM design node, distinct from finite-Fourier Q.

\begin{figure}[!htb]
\centering\includegraphics[width=\columnwidth]{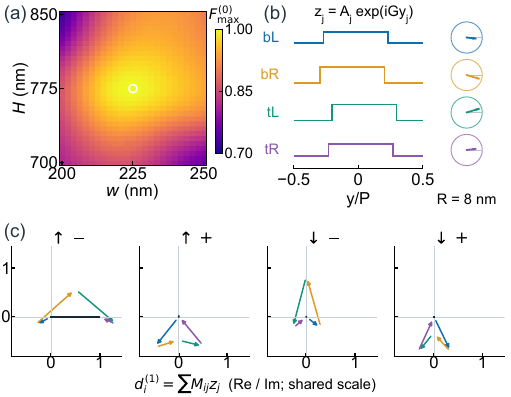}
\caption{The background bound and wall interference. (a) Circular-target reference bound $F_{\max}^{(0)}(\bm t)=F_{\max}(\bm t;C_0\mathsf P)$ over unmodulated rail cross sections. The circle marks the adopted parent. (b) Wall profiles and complex geometric inputs $z_j=A_j e^{\ii Gy_j}$. Arrow length and angle give $A_j$ and $Gy_j$, with a rightward zero-phase reference and 8-nm circle radii. (c) Their complex radiation contributions in the parent field. The wall contributions add in the selected channel and nearly cancel in the other three.}
\label{fig:mechanism}
\end{figure}

\begin{figure*}[!t]
\centering\includegraphics[width=\textwidth]{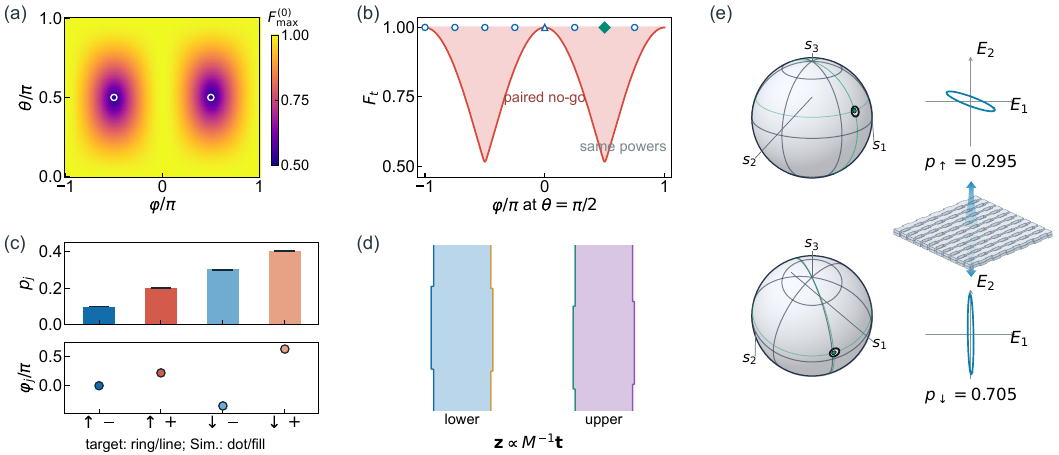}
\caption{Radiation states with paired and independent walls. (a) Single-side reference bound $F_{\max}^{(0)}$ versus power angle $\theta$ and relative phase $\varphi$. (b) Equal-power cut; shading marks states excluded by pairing, and finite markers use independent walls. (c) Target and FEM channel powers (top) and phases (bottom). (d) Corresponding wall masks. (e) Conditional polarizations and power shares. The central metasurface array is schematic. Ellipses share a common field-amplitude scale, and their arrows follow increasing time. Target powers and phases use horizontal marks and rings; FEM powers and phases use filled bars and dots. Target and FEM ellipses use dashed and solid lines, respectively.}
\label{fig:control}
\end{figure*}

To understand how this design realizes CURs, we resolve the radiation into contributions from the four modulated sidewalls. Each wall supplies a complex modulation harmonic $z_j=A_j e^{\ii Gy_j}$, where $G=2\pi/P$; its amplitude and longitudinal shift control the strength and phase of its radiation contribution. To first order, these contributions combine as $\bm d^{(1)}=M\bm z$, where each column of $M$ gives the radiation induced by one wall harmonic under a common energy normalization~\cite{johnson2002}. We select the paired modulations so that the contributions reinforce the upper negative-helicity channel and nearly cancel in the other three [Fig.~\ref{fig:mechanism}(b,c)].

However, the radiation amplitudes cannot be prescribed freely within an inversion-paired structure. Beyond the interference controlled by wall modulation, inversion symmetry and reciprocity impose a compatibility condition through the nonresonant background. We next derive a sharp fidelity bound that determines which target states this condition excludes. Let $C$ be the nonresonant scattering matrix and $\mathsf P_{ij}=\delta_{i,5-j}$ the inversion map in the stated channel order. At generic nonzero momentum, inversion relates distinct counterpropagating resonances. Reciprocity gives $C(\bm k)^T=C(-\bm k)$, and inversion gives $C(-\bm k)=\mathsf P C(\bm k)\mathsf P$. For a lossless background, $B=C\mathsf P$ is symmetric and unitary. Combining inversion with $C(\bm k)\bm d^*(-\bm k)=-\bm d(\bm k)$ gives $B\bm d^*=-\bm d$~\cite{hsu2017}. The fidelity to a unit target $\bm t$ has the sharp bound
\begin{equation}
 F_t=|\bm t^\dagger\bm n|^2\leq
 F_{\max}(\bm t;B)=\frac{1+|\bm t^\dagger B\bm t^*|}{2}.
 \label{eq:bound}
\end{equation}
A Takagi factorization $B=VV^T$ makes the compatible radiation space explicit: $\bm n=\ii V\bm x$, where $\bm x$ is a real unit vector. Optimization over this space therefore yields Eq.~\eqref{eq:bound}, attainable when the wall-coupling map reaches a maximizing state. The full derivation and equality condition are given in SM Sec.~S2.

For the symmetric, polarization-diagonal background considered here, the circular single-side target obeys $\beta_{\rm side,CP}\leq1/2+|t_x-t_y|/4$, where $t_x$ and $t_y$ are the linear-polarization transmission amplitudes. Nearly unit transmission amplitudes with a relative phase close to $\pi$ provide a nearly reflectionless half-wave background that permits near-perfect channel selection. We accordingly choose a parent cross section with a reference bound of approximately 0.9997 [Fig.~\ref{fig:mechanism}(a)], evaluated using the unmodulated background $C_0\approx C$ at weak modulation.

We further extend the theory from CURs to the joint geometric design of arbitrary coherent emission states and radiative lifetimes. This requires prescribing the relative amplitudes and phases across all four outgoing channels. However, selecting a single channel does not guarantee that arbitrary relative phases can be prescribed, even when the channel powers are unchanged. To expose this distinction, we consider the equal-power upper-side targets $\bm t=(1,e^{\ii\varphi},0,0)^T/\sqrt2$, which describe single-sided linearly polarized radiation with an orientation set by $\varphi$. As $\varphi$ changes from $0$ to $\pi/2$, the reference fidelity bound falls from nearly unity to 0.516 [Fig.~\ref{fig:control}(a,b)], although the channel powers remain fixed. Inversion pairing therefore excludes targets distinguished solely by their relative phase, revealing a restriction that chirality and directionality measures cannot detect because they retain only channel powers.

The same Takagi form also clarifies the origin of this restriction. At a fixed background, inversion pairing confines the radiation vector to four real coefficients, leaving three continuous state parameters after normalization; opposite signs represent the same physical state. Thus, additional geometric controls can explore this allowed space, but they cannot recover radiation states excluded by the background compatibility condition. Access to such states therefore calls for a change in the background or a release of inversion pairing.

We next remove the pairing constraint by modulating the four sidewalls independently. The wall harmonics $\bm z$ then provide four independent complex controls, and the coupling matrix $M$ has complex rank four with a condition number of about 8.25. The first-order map is therefore invertible, allowing a prescribed radiation vector to determine the wall modulation directly. Removing the common phase and magnitude leaves six parameters for the radiation state, while the magnitude sets the radiative lifetime. Thus, a normalized target $\bm t$ and a requested $Q_*$ jointly determine
\begin{equation}
 \bm z=\sqrt{\frac{\omega_0}{Q_*}}\,M^{-1}\bm t.
 \label{eq:inverse}
\end{equation}
The inverse prescription therefore assigns complementary roles to the wall modulations: relative amplitudes and shifts specify the coherent radiation state, while their common amplitude sets the radiative lifetime to first order. The corresponding inverse-design formulation is detailed in SM Sec.~S2. This control applies at fixed nonzero momentum and preserves reciprocity between opposite momenta; at a time-reversal-invariant momentum such as $\Gamma$, however, an isolated mode is its own partner and may obey an additional same-mode constraint even after inversion pairing is released.
% CHECK: The Gamma qualification is a scope statement; do not imply that SM S2 provides a dedicated Gamma-point derivation.

We next test whether the inverse-designed structures reproduce both the prescribed coherent radiation state and radiative lifetime. Figure~\ref{fig:control}(c--e) compares the channel amplitudes and phases of a general four-channel target with the calculated output, while four independent finite-element examples yield near-unity fidelities. Furthermore, finite vector-Fourier tests of four targets at $Q_*=4000$, 7000, and 10000 also yield near-unity fidelities, with the calculated $Q$ values 1.8\%--6.0\% above the prescribed values. The target geometries and state--lifetime comparisons are provided in SM Sec.~S3. Independent-wall designs for single-sided elliptically polarized targets are also verified in SM Sec.~S3. This extends geometric polarization and wavefront control~\cite{overvig2020,overvig2021,overvig2024,wangvbg2026} to coherent radiation across both half-spaces and its lifetime.
% CHECK: Table S1 supplies no Q_* for its FEM masks. Table S2 Q values systematically exceed Q_*; a common amplitude rescale has not been tested.

\begin{figure}[!htbp]
\centering\includegraphics[width=\columnwidth]{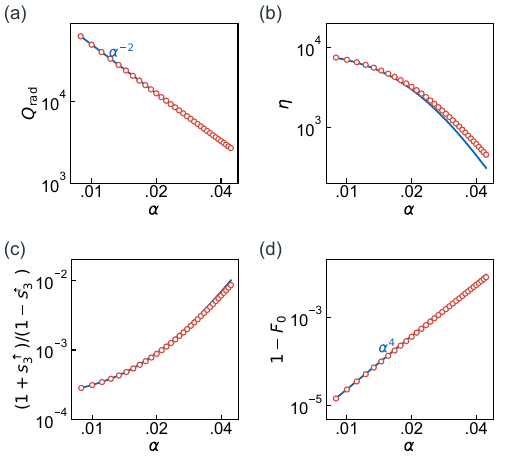}
\caption{Finite-modulation corrections at fixed relative wall amplitudes and shifts. (a) Radiative quality factor, (b) directionality ratio, (c) $(1+s_3^\up)/(1-s_3^\up)=P_{\up+}/P_{\up-}$, and (d) infidelity relative to the leading radiation state. Curves are predictions from the unmodulated operator. Open points are 33 finite-Fourier solutions. The fourth-order line in (d) has no fitted coefficient.}
\label{fig:rate}
\end{figure}

In addition, we establish the scaling laws governing radiative lifetime and radiation-state fidelity as the modulation strength varies. We scale every wall by $\alpha=a/w$ at fixed relative amplitudes and shifts. Sign reversal is equivalent to a half-period translation, so the resonance frequency is even in $\alpha$ and the radiation amplitude is odd in the corresponding phase convention~\cite{alpeggiani2017,sauvan2013,lalanne2018}.
\begin{equation}
 \widetilde{\bm c}(\alpha)=\alpha\bm g_1+\alpha^3\bm g_3+\cdots,
 \qquad \bm n=\frac{\widetilde{\bm c}}{\|\widetilde{\bm c}\|}.
 \label{eq:expansion}
\end{equation}
To leading order, common amplitude scaling changes the radiation strength but leaves the normalized state unchanged, so that $Q\propto\alpha^{-2}$. The CUR lifetime can therefore be increased arbitrarily by reducing the modulation strength in the lossless weak-modulation limit [Fig.~\ref{fig:rate}(a)], while the cubic radiation term governs the accompanying changes in channel ratios at finite modulation [Fig.~\ref{fig:rate}(b,c)]. The unmodulated parent operator determines both coefficients, linking lifetime tuning and state evolution within the same expansion. The expansion coefficients and their normalization are detailed in SM Sec.~S4.

\begin{figure}[!htbp]
\centering\includegraphics[width=\columnwidth]{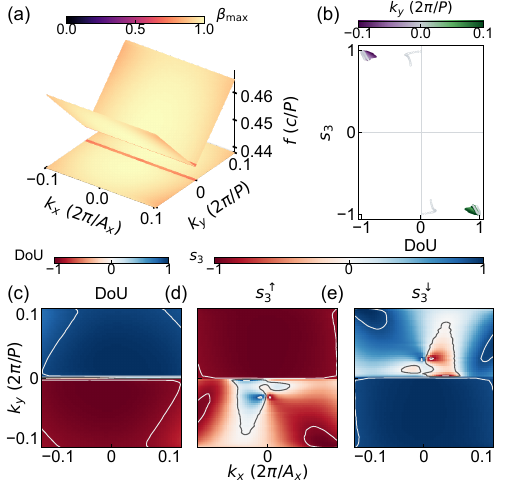}
\caption{Momentum persistence of the fixed CUR structure. (a) Upper-frequency FEM dispersion, colored by the largest single-channel fraction $\beta_{\max}$. (b) DoU versus the brighter side's $s_3$, colored by $K_y$. (c) Signed DoU. (d,e) Upper- and lower-side Stokes parameters. In (c--e), light contours mark $\pm0.9$ and dark contours mark zero. All panels use the complete $65\times65$ FEM grid; the dominant channel may vary with momentum. Here $K_x=k_xA_x/(2\pi)$ and $K_y=k_yP/(2\pi)$, with $A_x=850$~nm and $P\approx628.1$~nm. Full radiation statistics and branch comparisons are presented in SM Sec.~S5.}
\label{fig:domain}
\end{figure}

Interestingly, the radiation-state infidelity begins at order $\alpha^4$, while the radiative loss already scales as $\alpha^2$. To see why, we resolve the cubic radiation $\bm g_3$ into components parallel and perpendicular to the leading state $\bm n_0=\bm g_1/\|\bm g_1\|$. The parallel component rescales the radiation vector, whereas the perpendicular component rotates the normalized state by an amount of order $\alpha^2$. Consequently, the infidelity $1-F_0$, with $F_0=|\bm n_0^\dagger\bm n|^2$, begins at fourth order and takes the form
\begin{equation}
 1-F_0=\frac{\|(I-\bm n_0\bm n_0^\dagger)\bm g_3\|^2}{\|\bm g_1\|^2}\alpha^4+O(\alpha^6).
 \label{eq:quartic}
\end{equation}
The finite-Fourier solutions reproduce both the fourth-order scaling and its predicted coefficient [Fig.~\ref{fig:rate}(d)], with a logarithmic slope of 3.96 at $a=2$--4~nm. These contrasting scalings therefore permit substantial lifetime tuning with only small changes in the radiation state. SM Sec.~S4 details the coefficient and finite-modulation comparisons.

Finally, we examine the momentum robustness of the designed CUR emission. Figure~\ref{fig:domain} maps the dispersion, directionality, and half-space Stokes parameters, demonstrating persistent channel selection over a finite momentum region~\cite{zhen2014,doeleman2018,jin2019}. The largest single-channel fraction $\beta_{\max}=\max_jP_j/P_{\rm rad}$ directly quantifies this joint selection [Fig.~\ref{fig:domain}(a)]. Away from the narrow counterpropagating-parent mixing region, the mode becomes predominantly traveling and retains a dominant circularly polarized output: for $|K_y|\geq0.00625$ on the sampled upper branch, the brighter half-space carries at least 87.3\% of the power, with an opposite-to-selected helicity power ratio below 6.45\%. Near the axis, however, coherent mixing of the $+G$ and $-G$ guided parents changes the interference responsible for single-sided radiation, even when each parent contribution is nearly circular [Fig.~\ref{fig:realization}(c)]~\cite{feshbach1958,li1996}. The near-axis mixing and full momentum-resolved radiation statistics are detailed in SM Sec.~S5.

The weak-half-space radiation also reveals contributions beyond the guided-parent subspace, whose interference with the parent radiation can appreciably change the polarization. The corresponding field decomposition is presented in SM Sec.~S5.4. The dominance of one traveling-parent harmonic, together with a slowly varying wall-coupling map, allows the designed interference to persist as momentum changes, thereby sustaining joint directional and helicity selection beyond the design point. The supporting parent-harmonic and radiation analyses are presented in SM Sec.~S5.

Our demonstrations span CURs, single-sided elliptically polarized states, and general four-channel superpositions, connecting selective emission with full coherent radiation control. Here, CUR selectivity concerns one eigenmode at fixed momentum, distinct from angular shaping of chiral emission~\cite{gromyko2024}. Looking ahead, CURs with tunable radiative lifetimes could provide resonant building blocks for single-sided circularly polarized light sources~\cite{kodigala2017,dixon2021}. More broadly, combining coherent radiation control with spontaneous-emission engineering could allow the design of how emitters radiate into selected optical channels~\cite{wangse2026}. Through reciprocity, the same framework also offers a route to matching resonances with prescribed coherent incident fields~\cite{arbabi2015,rubin2021,wanginterconnect2024}. Joint control of the radiation state and lifetime therefore provides a common design basis for light emission and optical coupling.

In summary, we have introduced and realized CURs with arbitrarily high radiative quality factors and extended their design to arbitrary coherent emission states with prescribed lifetimes. Our unified theory identifies the phase constraints imposed by inversion pairing and provides an invertible first-order geometric prescription when the sidewalls are independent. In addition, the scaling laws explain how lifetime tuning preserves the leading radiation state, while momentum maps demonstrate persistent channel selection beyond the design point. Complete coherent radiation control therefore extends resonance engineering from selecting particular channels to shaping their full complex superposition together with the radiative lifetime.
\label{end:body}
\begin{acknowledgments}
This work was supported by the Natural Science Foundation of Sichuan Province (Grant No. 2024NSFSC0460).
\end{acknowledgments}
\bibliographystyle{apsrev4-2}
\bibliography{ChiralUGR-PRL-References}
\end{document}